\documentclass[aps,prl,twocolumn,superscriptaddress,showpacs,preprintnumbers,amsmath,amssymb]{revtex4}
\usepackage{epsfig}
\usepackage{graphicx} 
\graphicspath{{ps}}

{\renewcommand{\thefootnote}{\fnsymbol{footnote}}
\newcommand{\mb}{\ensuremath{M_{\mathrm{bc}}}}

\begin{document}
\preprint{\vbox{ \hbox{ }
                 \hbox{ }
                 \hbox{ }
}}

\title{
 \quad\\[0.5cm] 
\boldmath
Measurement of the Branching Fractions for $B \to \omega K$ and
$B \to \omega \pi$.}
%
\affiliation{Budker Institute of Nuclear Physics, Novosibirsk}
\affiliation{Chiba University, Chiba}
\affiliation{University of Cincinnati, Cincinnati, Ohio 45221}
\affiliation{Gyeongsang National University, Chinju}
\affiliation{University of Hawaii, Honolulu, Hawaii 96822}
\affiliation{High Energy Accelerator Research Organization (KEK), Tsukuba}
\affiliation{Hiroshima Institute of Technology, Hiroshima}
\affiliation{Institute of High Energy Physics, Chinese Academy of Sciences, Beijing}
\affiliation{Institute of High Energy Physics, Vienna}
\affiliation{Institute for Theoretical and Experimental Physics, Moscow}
\affiliation{J. Stefan Institute, Ljubljana}
\affiliation{Kanagawa University, Yokohama}
\affiliation{Korea University, Seoul}
\affiliation{Kyungpook National University, Taegu}
\affiliation{Swiss Federal Institute of Technology of Lausanne, EPFL, Lausanne}
\affiliation{University of Ljubljana, Ljubljana}
\affiliation{University of Maribor, Maribor}
\affiliation{University of Melbourne, Victoria}
\affiliation{Nagoya University, Nagoya}
\affiliation{Nara Women's University, Nara}
\affiliation{National United University, Miao Li}
\affiliation{Department of Physics, National Taiwan University, Taipei}
\affiliation{H. Niewodniczanski Institute of Nuclear Physics, Krakow}
\affiliation{Nihon Dental College, Niigata}
\affiliation{Niigata University, Niigata}
\affiliation{Osaka City University, Osaka}
\affiliation{Osaka University, Osaka}
\affiliation{Panjab University, Chandigarh}
\affiliation{Peking University, Beijing}
\affiliation{Princeton University, Princeton, New Jersey 08545}
\affiliation{University of Science and Technology of China, Hefei}
\affiliation{Seoul National University, Seoul}
\affiliation{Sungkyunkwan University, Suwon}
\affiliation{University of Sydney, Sydney NSW}
\affiliation{Tata Institute of Fundamental Research, Bombay}
\affiliation{Toho University, Funabashi}
\affiliation{Tohoku Gakuin University, Tagajo}
\affiliation{Tohoku University, Sendai}
\affiliation{Department of Physics, University of Tokyo, Tokyo}
\affiliation{Tokyo Institute of Technology, Tokyo}
\affiliation{Tokyo Metropolitan University, Tokyo}
\affiliation{Tokyo University of Agriculture and Technology, Tokyo}
\affiliation{Toyama National College of Maritime Technology, Toyama}
\affiliation{University of Tsukuba, Tsukuba}
\affiliation{Virginia Polytechnic Institute and State University, Blacksburg, Virginia 24061}
\affiliation{Yokkaichi University, Yokkaichi}
\affiliation{Yonsei University, Seoul}
  \author{C.~H.~Wang}\affiliation{National United University, Miao Li} 
  \author{K.~Abe}\affiliation{High Energy Accelerator Research Organization (KEK), Tsukuba} 
  \author{K.~Abe}\affiliation{Tohoku Gakuin University, Tagajo} 
  \author{N.~Abe}\affiliation{Tokyo Institute of Technology, Tokyo} 
  \author{T.~Abe}\affiliation{High Energy Accelerator Research Organization (KEK), Tsukuba} 
  \author{I.~Adachi}\affiliation{High Energy Accelerator Research Organization (KEK), Tsukuba} 
  \author{H.~Aihara}\affiliation{Department of Physics, University of Tokyo, Tokyo} 
  \author{M.~Akatsu}\affiliation{Nagoya University, Nagoya} 
  \author{T.~Aso}\affiliation{Toyama National College of Maritime Technology, Toyama} 
  \author{V.~Aulchenko}\affiliation{Budker Institute of Nuclear Physics, Novosibirsk} 
% ...
\collaboration{The Belle Collaboration}

\noaffiliation

\begin{abstract}
We report improved measurements of branching fractions for
charmless hadronic two-body {\it B} meson decays containing
an $\omega$ meson in the final state. The results are 
based on a data sample of 78 fb$^{-1}$ collected on 
the $\Upsilon(4S)$ resonance by the Belle
detector. We measure the branching fractions ${\mathcal B}(B^+
\to \omega K^+) = (6.5^{+1.3}_{-1.2}\pm 0.6)\times 10^{-6}$ and
${\mathcal B}(B^+ \to \omega \pi^+) = (5.7^{+1.4}_{-1.3}\pm
0.6)\times 10^{-6}.$ We give 90\% confidence upper limits for
${\mathcal B}(B^0 \to \omega K^0) < 7.6\times 10^{-6}$ and
${\mathcal B}(B^0 \to \omega \pi^0) < 1.9\times 10^{-6}.$ We also
obtain the partial rate asymmetries ${\mathcal
A}_{CP}=0.06^{+0.21}_{-0.18}\pm 0.01$ for $B^\pm \to \omega K^\pm$
and ${\mathcal A}_{CP}=0.50^{+0.23}_{-0.20}\pm 0.02$ for $B^\pm
\to \omega \pi^\pm.$
\end{abstract}

\maketitle

\tighten
{\renewcommand{\thefootnote}{\fnsymbol{footnote}}}
\setcounter{footnote}{0}
\section{Introduction}
Charmless hadronic $B$ decays play an important role in the
understanding of CP violation in the $B$ system. These decays
proceed primarily through interfering $b\to s$ loop penguin diagrams and $b
\to u$ tree spectator diagrams. 
Studies of $B \to \omega h$, where $h$ denotes $K^+$, $\pi^+$, 
$K^0$, and $\pi^0$, are important examples of such 
decays. Charge conjugates are implied unless otherwise stated.
We also assume equal production of $B^+B^-$ and $B^0\bar{B}^0$
pairs from the $\Upsilon(4S)$.

\begin{table}[htb]
\begin{center}
\caption{Measurements of branching fractions for $B^+ \to \omega
K^+$ and $B^+ \to \omega \pi^+$ from CLEO, BaBar and Belle. The
units are $10^{-6}$.}
\begin{tabular}{l|c|c|c} \hline\hline
Mode & CLEO\cite{CLEO2} &
BaBar\cite{BaBar1} & Belle\cite{Belle} \\ \hline
$\omega K^+$
& $3.2^{+2.4}_{-1.9}\pm 0.8\hspace{0em}$
& $5.5\pm 0.9 \pm 0.5\hspace{0em}$
& $9.2^{+2.6}_{-2.3}\pm{1.0}$ \\
$\omega \pi^+$
& $11.3^{+3.3}_{-2.9}\pm{1.4}$ & $4.8\pm 0.8 \pm{0.4}$
& $4.2^{+2.0}_{-1.8}\pm{0.5}\hspace{0em}$
  \\\hline \hline
\end{tabular}
\label{old}
\end{center}
\end{table}
Table~\ref{old} lists the branching fractions from previous
measurements\cite{CLEO1,CLEO2,BaBar,BaBar1,Belle}, which indicate 
some discrepancies for $B^+ \to \omega K^+$.
Naive factorization and QCD factorization
approaches~\cite{theory, theory1}
yield values of ${\mathcal B}(B^+ \to \omega \pi^+)$ consistent
with the experimental results. However, these approaches predict
${\mathcal B}(B^+ \to \omega \pi^+)$ to be a factor of two larger
than ${\mathcal B}(B^+ \to \omega K^+)$, which is not supported by
Belle's previous experimental results that were based on a $29.4$ fb$^{-1}$ 
data sample~\cite{Belle}.
In this paper, we update our previous measurements 
on $\omega K^+$ and $\omega \pi^+$
with a 78.1 fb$^{-1}$ data sample. We also
report measurements of $\omega K^0$ and $\omega \pi^0$
decay modes.
\section{Apparatus and Data Set}
The data sample used was collected with the Belle detector at the
KEKB asymmetric energy $e^+e^-$ collider~\cite{kekb}, which
collides 8.0 GeV $e^-$ and 3.5 GeV $e^+$ beams at 
a small crossing angle ($\pm 11$ mrad). 
The data sample contained $85.0\times 10^6$ $B\bar{B}$ pairs produced
at the $\Upsilon(4S)$ resonance. A 8.8 fb$^{-1}$ data sample
taken at a center-of-mass energy 60 MeV below the
$\Upsilon(4S)$ is used to characterize continuum background.
In order to establish the event selection criteria, we use a
Monte Carlo (MC) generator~\cite{qq} to generate signal, generic
$b \rightarrow c$, and other charmless rare {\it B} decays.
The GEANT3 package~\cite{geant} is used for detector simulation.

The Belle detector measures charged particles and
photons with high efficiency and precision\cite{NIM}.
Charged particle tracking is provided
by a silicon vertex detector (SVD) and a central drift chamber
(CDC) that surround the interaction region. The charged particle
acceptance covers the laboratory polar angle region between $\theta=17^o$
and $150^o$. Charged hadrons are distinguished by combining the
responses from an array of silica aerogel \v{C}erenkov counters
(ACC), a barrel-like array of 128 time-of-flight scintillation
counters (TOF), and $dE/dx$ measurements in the CDC. The combined
response provides $K/\pi$ separation of at least $2.5\sigma$ for
laboratory momentum up to 3.5 GeV/$c$. Electromagnetic showers are
detected in an array of 8736 CsI(Tl) crystals (ECL) located inside
the magnetic volume, which covers the same solid angle as the
charged particle tracking system. 
The magnet return yoke consists of alternating layers 
of resistive plate counters and 4.7 cm thick steel plates for
detecting $K^0_L$'s and identifying muons.
%
\section{Event Selection}
Hadronic events are selected using criteria based on the charged track
multiplicity and total visible energy sum; the efficiency is
greater than 99\% for generic $B\bar B$ events~\cite{trig}. All primary
charged tracks must satisfy quality
requirements based on their impact parameters relative 
to the run-dependent interaction point (IP). 
The deviation from the IP position is required
to be within $\pm 1.5$ cm in the transverse direction and $\pm 2$
cm in the longitudinal direction. 
Charged particle identification
is based on the ratio KID$={\mathcal L_K}$/(${\mathcal L_\pi+ \mathcal L_K}$),
where ${\mathcal L}_K$ and ${\mathcal L}_{\pi}$ are 
likelihoods for $K$ and $\pi$ hypotheses. 
A higher value of KID indicates a more kaon-like particle.
$\pi^0$ meson candidates are reconstructed from pairs of photons,
each consisting of energy clusters greater than 50~MeV, with
$\gamma\gamma$ invariant mass within $3\sigma$ ($\sigma=5.4$
MeV/$c^2$) of the $\pi^0$ mass. $K^0_S$ meson candidates
are reconstructed using pairs of oppositely charged particles
that have an invariant mass in the range
480 MeV/$c^2< m(\pi^+\pi^-) <$ 516 MeV/$c^2$.
The vertex of the $K^0_S$ candidate is required to be well
reconstructed and displaced from the interaction point, and the
$K^0_S$ momentum direction must be consistent with the $K^0_S$
flight direction. 
Candidate $\omega \to \pi^+\pi^-\pi^0$ decays are reconstructed
from charged pions with KID $< 0.9$ and $\pi^0$s with center-of-mass 
frame momentum greater than 0.35~GeV/$c$. 
The $\omega$ meson candidates are required 
to have an invariant mass within $\pm 30$ MeV/$c^2$ of the nominal 
value $(\pm 2 \sigma)$.
\section{$B$ Reconstruction}
$B$ meson candidates are formed by combining an $\omega$ meson
with either a kaon ($K^+$, $K^0$) or a pion ($\pi^+$, $\pi^0$). 
We require KID $>0.6$ and KID $<0.4$ for $K^+$ and $\pi^+$,
respectively. Studies from
$D^{*+}\to D^0 \pi^+(D^0 \to K^-\pi^+)$ decays give
particle identification efficiencies,
$\epsilon_K = 85\%$ and $\epsilon_\pi =89\%$ with 
misidentification rates, $f_{\pi}= 8\%$ and 
$f_{K}= 11\%$, respectively.

$B$ meson candidates are then identified using
the beam constrained mass $\mb$ = $\sqrt{(E^{\rm CM}_{\rm
beam})^2-|P^{\rm CM}_B|^2}$ and the energy difference $\Delta E =
E^{\rm CM}_B - E^{\rm CM}_{\rm beam}$, where $E^{\rm CM}_{\rm
beam} = 5.29$~GeV, and $P^{\rm CM}_B$, $E^{\rm CM}_B$ are the
momentum and energy of the $B$ candidate in the $\Upsilon(4S)$ rest
frame. For the $\Delta E$ calculation, the kaon in candidate 
$B^+ \to \omega K^+$ decays is assigned
a pion mass so that $\omega K^+$ and $\omega \pi^+$
can be fit simultaneously.
For events with multiple candidates, the best candidate is 
selected using the quality of the $B$ vertex fit.
According to signal MC, 
the resolutions for $\mb$ and $\Delta
E$ are 3~MeV/$c^2$ and 24~MeV respectively  
for $B \to \omega K^+$, $\omega \pi^+$ and $\omega K^0$ decays. 
For the decay $B \to \omega \pi^0$, the resolutions are 3.5~MeV/$c^2$
for $\mb$ and 55~MeV for $\Delta E$.

The $B$ candidates are required to be within the rectangular region in the
$\mb$ - $\Delta E$ plane, $5.2$~GeV/$c^2$$< \mb < 5.3$~GeV/$c^2$ and
$|\Delta E|<0.25$~GeV. Signal regions of
$5.27$~GeV/$c^2< \mb < 5.3$~GeV/$c^2$ and $|\Delta E|<0.10$~GeV are used
to display fit projections. Sideband
regions are defined 
as $5.2$~GeV/$c^2 < \mb < 5.26$~GeV/$c^2$ with $|\Delta
E|<0.25$~GeV for $\Delta E$, and
$5.2$~GeV/$c^2 < \mb $ with $0.10$~GeV $<|\Delta
E|<0.25$~GeV for $\mb$.

Since $B \to \omega h$ is a $P \to$ $V$ $P$ decay,
where $V$ means vector and $P$ means pseudo-scalar particles, 
the $\omega$ meson is polarized. The
$\omega$ helicity angle, $\theta_{\rm hel}$, is defined as the angle
between the $B$ flight direction and the vector perpendicular to the
$\omega$ decay plane in the $\omega$ rest frame.
Further background suppression is achieved using the $\omega$ helicity and
the quality of the $B$ vertex fit $(\chi^2_B)$.
%
%
\section{Background suppression}
Backgrounds from $b \to c$ decays and the feed-across from other
charmless rare $B$ decays are found to be negligible using MC 
simulations that assume the best known branching fraction for each decay. 
The dominant backgrounds arise from
the $e^+e^- \to q \bar{q}$ ($q = u$, $d$, $s$ or $c$)
continuum process, which has a jet-like event topology in contrast to 
the spherical $B\bar{B}$ events.

Several event-shape variables are used to distinguish between
$B$ decays and continuum background. The thrust angle $\theta_T$
is defined as the angle between the primary $B$ decay daughter
$\omega$ and the thrust axis formed by all the particles from
the other $B$. $S_\perp$ is the scalar sum of the
transverse momenta of all particles outside a $45^{\circ}$ cone
around the primary $B$ decay daughter direction divided by the
scalar sum of their momenta. In addition to these, a set of
variables derived from Fox-Wolfram moments~\cite{fw} are used. The
moments are defined by
$$ R^{so}_l = {{\sum_{i,k}|p_i||p_k|P_l(\cos\theta_{ik})}\over
   {\sum_{i,k}|p_i||p_k|}},$$ and
$$ R^{oo}_l = {{\sum_{i,j}|p_i||p_j|P_l(\cos\theta_{ij})}
   \over{\sum_{i,j}|p_i||p_j|}},$$
where $p$ stands for particle momentum, and $P_l$ is the $l^{\rm th}$ 
Legendre polynomial. There are two groups of particles that go into this
summation. The index $k$ refers to (neutral or charged) particles from the $B$
candidate, while $i$ and $j$ refer to other particles not from
that $B$ candidate. $R^{so}_1$, $R^{so}_3$ and $R^{oo}_1$ are not
used because of their strong correlation with $\mb$. To optimize the
discrimination, the remaining 5 variables ($l \le 4$) are combined
with $\cos\theta_T$ and $S_\perp$ to form a Fisher discriminant
${\mathcal F}$~\cite{fisher,etapk}. The cosine of the angle
between the $B$ flight direction and the beam axis
($\cos\theta_B$), and ${\mathcal F}$ are found to be independent,
and their probability density functions (PDFs) are obtained by
using MC samples for signal, and off-resonance data for continuum
background. The variables $\cos\theta_B$ and ${\mathcal F}$ are
then combined to form a likelihood ratio ${\mathcal LR} = {\mathcal
L}_s/({\mathcal L}_s + {\mathcal L}_{bg})$, where ${\mathcal
L}_{s(bg)}$ is the product of signal ($q\bar q$) PDFs. 
A selection requirement is imposed on $\mathcal LR$ to reject
continuum background.
A typical cut is ${\mathcal LR} > 0.5$ and retains approximately
$83 \%$ of the signal candidates while reducing the 
background by approximately $73 \%$.
\section{Analysis}
Signal yields are obtained using $\mb$ and
$\Delta E$ as independent variables in an extended unbinned
maximum likelihood (ML) fit after restrictions are imposed
on the variables $\chi^2_B$,
${\mathcal LR}$ and $\cos\theta_{\rm hel}$. 
These 
are: ${\mathcal LR} > 0.65$ for $\omega K^+/\pi^+$, ${\mathcal LR}
> 0.5$ for $\omega K^0$, ${\mathcal LR} > 0.8$ for $\omega \pi^0$
and $|\cos\theta_{\rm hel}|>0.5$. For $N$ input
candidates, the likelihood is
defined as
\begin{center}
$
{\mathcal L}(N_S,N_B) = {e^{-(N_S+N_B)}} \prod_{i=1}^{N}
[N_{S} P_{S_i}(\mb)P_{S_i}(\Delta E) + 
N_{B} P_{B_i}(\mb)P_{B_i}(\Delta E)],
$
\end{center}
where the index $i$ runs over each event,
$P_{S_i}$ and $P_{B_i}$ are the probability densities as
functions of $\mb$ and $\Delta E$ for signal
and background, respectively. This method treats the extracted
yields for signal $N_S$ and background $N_B$ according to 
Poisson statistics
and constrains their sum to the observed number of candidates $N$
at the maximum likelihood.

The signal PDFs are determined from signal MC 
while the continuum background PDFs are derived from 
the off-resonance data. The background shapes are
verified using data from the sideband region. The PDF 
for the $\Delta E$ background is modeled by a second-order 
polynomial function. 
The PDF for the $\mb$ background distribution is modeled with 
a smooth function with
parameters determined from off-resonance data~\cite{argus}. To
model the low energy tail, the $\Delta E$ signal PDFs use
a \lq\lq Crystal Ball" line shape function~\cite{CBLINE} with parameters
determined by fits to signal MC. The $\mb$ PDFs are
the sum of two Gaussian functions with different widths, which were
obtained by fits to signal MC. Studies of $B^+ \to {\bar D^0}
\pi^+$ and ${\bar D^0} \to K^+ \pi^-\pi^0$ decays were used to fix the
mean $\mb$. Differences between widths obtained in these
studies and those from the signal MC are regarded as
systematic uncertainties.
\begin{figure}
\includegraphics[width=0.52\textwidth]{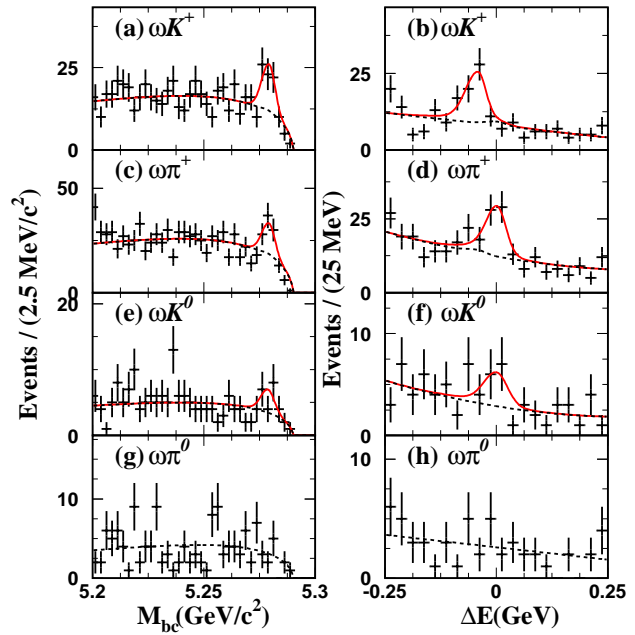}
\caption{Signal region projections of $\mb$ (left) and 
$\Delta E$ (right) for $\omega K^+$, $\omega \pi^+$, $\omega K^0$ and
$\omega \pi^0$. The solid curves show the results
of the 2D fits with the background components represented
as dashed curves. Small background enhancements near $0$
MeV in (b) and $-50$ MeV in (d) are from misidentified
$B^+ \to \omega \pi^+$ and $B^+ \to \omega K^+$ decays.}
\label{omefitmb}
\end{figure}
%
%

The overall reconstruction efficiencies, $\epsilon$, 
are the products of detection efficiencies,
determined from MC with no KID requirements,
and KID efficiencies determined from
$D^{*+}\to D^0 \pi^+$, $D^0 \to K^-\pi^+$ events in the data.
The statistical significance ($\Sigma$) is
defined as $\sqrt{-2{\ln}[{\mathcal L(0)}/{\mathcal L}_{\rm max}]}$,
where ${\mathcal L}_{\rm max}$ is the maximum likelihood at the
nominal signal yield and ${\mathcal L(0)}$ is the likelihood with
the signal fixed at zero. The 90\% confidence level upper limit is calculated
from the equation
$ \frac{\int_0^{x_{\rm max}} {\mathcal L}(x) \,dx}{\int_0^\infty {\mathcal L}(x) \,dx}
 = 90\% ,$
where only the statistical uncertainties are considered. For the final
upper limit, the above limit is increased by one standard
deviation of the systematic error.
\begin{table}[htb]
\begin{center}
\caption{Signal yields($N_s$), efficiencies($\epsilon_{tot}$) including
secondary decay branching fractions, fit
significances($\Sigma$), branching fractions(${\mathcal B}$),
and $90\%$ confidence level upper limits (UL)
on the branching fractions for $\omega K^0$ and $\omega \pi^0$.}
\begin{tabular}{lccccc} \hline\hline
Mode & $N_s$ & $\epsilon_{tot}(\%)$
     & $\Sigma$  & $\mathcal B(\times 10^{-6}) $  & UL$(\times 10^{-6})$ \\\hline
$\omega K^+$ & $44.6^{+9.1}_{-8.3}$  & 8.1
& $7.8 \sigma$ & $6.5^{+1.3}_{-1.2}\pm 0.6$  & -\\
$\omega \pi^+$ & $42.1^{+10.1}_{-9.3}$ & 8.7 & $6.0 \sigma$ &
$5.7^{+1.4}_{-1.3}\pm 0.6$ & -\\
$\omega K^0$ & $11.1^{+5.2}_{-4.4}$ & 3.3 & $3.2 \sigma$ &
$4.0^{+1.9}_{-1.6}\pm {0.5}$ & 7.6\\
$\omega \pi^0$ & $0^{+2.1}_{-0.0}$ & 5.2 &
 - & - & 1.9 \\
\hline\hline
\end{tabular}
\label{result}
\end{center}
\end{table}
\section{Measurements of Branching Fractions}
The results from the fits are shown in
Table~\ref{result}. 
Figure~\ref{omefitmb} shows the $\mb$ and $\Delta E$
distributions, where events in the $\mb$ ($\Delta E$) plots are
required to be in the $\Delta E$ $(\mb)$ signal region after
all selection criteria. The signal yields from the fits are
$N_{\omega K^+} = 44.6^{+9.1}_{-8.3}$, $N_{\omega \pi^+} =
42.1^{+10.1}_{-9.3}$ and 
$N_{\omega K^0} = 11.1^{+5.2}_{-4.4}$ (statistical errors only).
No signal is observed for $B^0 \to \omega \pi^0$.
For $B^+ \to \omega \pi^+$ and $\omega K^+$, the $\pi^+/K^+$
feed-across is not negligible and its level is fixed in the ML fit. 
For $B^+ \to \omega \pi^+$, the contribution is estimated by using
$\omega K^+$ yields from the fitted $B^+ \to \omega K^+$ 
candidate events assuming no feed-across from $\omega \pi^+$,
dividing by the kaon efficiency 
and multiplying by the kaon mis-identification probability.
The result is $4.8\pm 1.0$ events 
from $B^+ \to \omega K^+$ in the 
$B^+ \to \omega \pi^+ $ signal.
\begin{figure}
\includegraphics[width=0.50\textwidth]{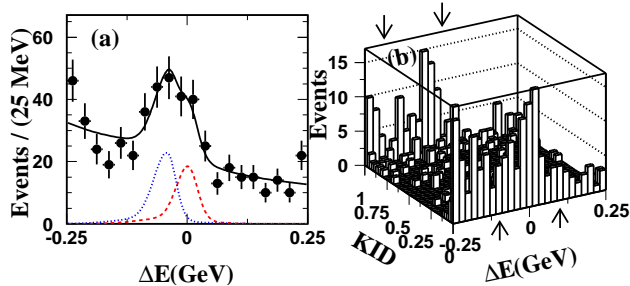}
\caption{ a): $\Delta E$ distributions without 
KID requirements in the $\mb$ signal region.
Dotted(dashed) curves indicate the signal components
$\omega K^+$ ($\omega \pi^+$) obtained from the
$\Delta E$ fit to the $B^+ \to \omega h^+$ candidate
events. 
The solid curve shows the sum of signal and continuum background
components.
b): KID $vs$ $\Delta E$ distributions in
the $\mb$ signal region. 
Arrows show the signal $\Delta E$ region with
$KID>0.6$ for $\omega K^+$ and $KID<0.4$ for $\omega \pi^+$.}
\label{dekidfig}
\end{figure}
This value is consistent with the level determined by repeating
the fit for $B^+ \to \omega \pi^+$ with 
the level of $\omega K^+$ feed-across left as a free 
parameter: $11.6 \pm 9.0$ events. 
This difference is assigned to the systematic
error of $\omega K^+$ feed-across for $\omega \pi^+$ decay.
A similar procedure is used to determine 
the $\omega \pi^+$ contamination in the 
$B^+ \to \omega K^+$ yield. Here
the feed-across is found to be $3.3\pm 0.7$ events.
The final measurements of branching fractions are listed in
Table~\ref{result}.

Because of the assignment of the pion mass to the kaon, the 
$B^+ \to \omega K^+$ signal peaks at $\Delta E = -50$ MeV,
which provides some discrimination from $B^+ \to \omega \pi^+$
events, which peak at $\Delta E = 0$ MeV. We use this to provide 
a consistency check of the $\omega K^+$ and $\omega \pi^+$ yields
by fitting to the $\Delta E$
distribution for $B^+ \to \omega h^+$ candidates with no 
KID requirements applied.
The signal yields are $60.0^{+15.5}_{-14.8}$ and $47.7^{+14.6}_{-13.7}$
events for $\omega K^+$ and $\omega \pi^+$, respectively, which
are consistent with the results using the KID, where 
the efficiency-corrected yields 
are $52.4^{+10.7}_{-9.8}$ events for $\omega K^+$ and $47.3^{+11.3}_{-11.0}$
events for $\omega \pi^+$.
Figure~\ref{dekidfig} shows the results of this fit 
and a lego plot of KID versus $\Delta E$. 
From the lego plot, there is a clear separation in the KID distribution 
between $\omega K^+$ and $\omega \pi^+$ signal yields, which also provides
a consistency check.

\begin{figure}
\includegraphics[width=0.48\textwidth]{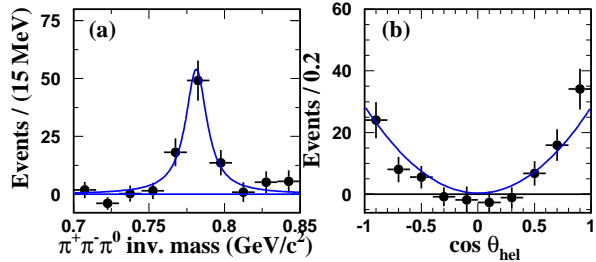}
\caption{Fitted yields in bins of (a) $\pi^+\pi^-\pi^0$
invariant mass and (b) cosine of $\omega$ helicity angle for
$\omega K^+$ and $\omega \pi^+$. Solid curves show the distribution
from signal MC normalized to the results
from the fits.} \label{omekpihelfig}
\end{figure}

We also examine the properties of $\omega$
candidates in our fit sample. The clear $\omega$ mass peak and
polarized $\cos\theta_{\rm hel}$ distribution shown in
Fig.~\ref{omekpihelfig}(a) and (b) confirm our fitted signals are from
$\omega$ mesons with no significant non-resonant $\pi^+\pi^-\pi^0$
contribution. Several other consistency checks have also been
performed including tightening ${\mathcal LR}$ requirements, 
and performing 1-D ML fits to $\mb$ and $\Delta E$. 
All studies yield consistent results.

Systematic uncertainties for each mode are presented in
Table~\ref{system}, with contributions arising from the background
suppression, reconstruction and the fitting function variations.
The systematic errors from the MC modeling of the $\chi^2_B$ and
${\mathcal LR}$ requirements are studied with 
$B^+ \to {\bar D^0} \pi^+$ and ${\bar D^0} \to K^+ \pi^- \pi^0$ decays,
which are $2.0\%$ for $\mathcal LR$ and $3.0\%$ for $\chi^2_B$.  
We study the systematic error associated with the
$\omega$ polarization ($\cos\theta_{\rm hel}$) 
requirement $|\cos\theta_{\rm hel}|<0.5$ 
by comparing the MC distributions 
with the fitted yields distribution(Fig.~\ref{omekpihelfig}).
We assign a $2.6\%$ systematic error.
\begin{table}[htb]
\caption{Systematic errors for $\omega h$. Feed-across means
$\omega K^+(\pi^+)$ for $\omega \pi^+(K^+)$ decay.
The unit is in percent $(\%)$.}
\begin{tabular}{lcccc} \hline\hline
Mode & $\omega K^+$ & $\omega \pi^+$ & $\omega K^0$ &
$\omega \pi^0$ \\\hline
Background suppression   & 4.4  & 4.4 & 4.4 & 4.4 \\
Reconstruction  & 8.0  & 8.0 & 9.5 & 9.4 \\
Fit   & $^{+1.3}_{-1.6}$  & $^{+1.6}_{-1.8}$
 & $^{+4.4}_{-4.3}$ & - \\
Feed-across  & $\pm 1.6$ & $^{+3.6}_{-3.3}$ & - & - \\
$N_{B\bar B}$ & 1 & 1 & 1 & 1 \\
${\mathcal B}(\omega\to \pi^+\pi^-\pi^0)$  
& 0.8 & 0.8 & 0.8 & 0.8 \\\hline 
Sum & $9.4$ & $10.0$
& $11.4$ & $10.6$ \\\hline\hline
\end{tabular}
\label{system}
\end{table}
The total systematic error from background suppression is $4.4\%$. 
The systematic error due to uncertainties in the reconstruction
is determined from detailed studies of the charged particle
tracking, $\omega$ mass resolution, KID and $\pi^0$ detection.
For charged tracking and $\pi^0$ detection, the decay modes 
$\eta \to \gamma\gamma,\pi^+\pi^-\pi^0$ and $\eta \to \pi^0\pi^0\pi^0$ 
are used. 
By comparing results in data and MC,
we assign a relative error of $2.0\%$ for charged track reconstruction, 
$3.0\%$ for $\omega$ mass cut, $4\%$ for $\pi^0$ detection.
The total reconstruction systematic error ranges between
$8.0\%$ and $9.5\%$ . The mean and width
differences of $\mb$ and $\Delta E$ distributions between data and
MC from $B^+ \to {\bar D^0} \pi^+$ decays are included in the
systematic errors from fitting.
The systematic uncertainty on the branching fraction of 
$\omega \to \pi^+\pi^-\pi^0$ is 
obtained from the PDG tables~\cite{PDG}.
\section{$\mathcal A_{CP}$ Measurements}
\begin{figure}
\includegraphics[width=0.52\textwidth]{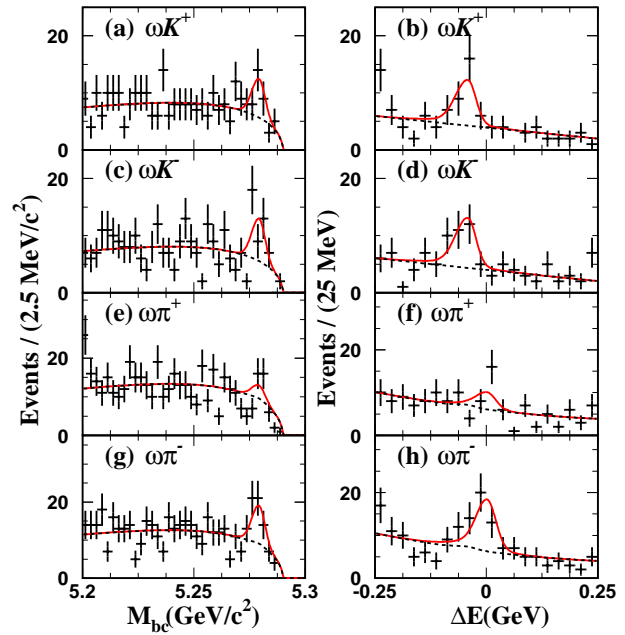}
\caption{Projections of the 2D
$\mb$ and $\Delta E$ for $B^\pm \to \omega K^\pm$
and $B^\pm \to \omega \pi^\pm$ decays.
Solid curves show the fit results.
The dashed curves indicate the backgrounds.}
\label{acpetap}
\end{figure}

We determine partial rate asymmetries
defined as
$$\mathcal A_{CP}
= {{N(B^- \to \omega h^-) - N(B^+ \to \omega h^+)}\over 
{N(B^- \to \omega h^-)+ N(B^+ \to \omega h^+)}}.$$ 
The values of $\mathcal A_{CP}$ were measured for 
the modes $B^\pm \to \omega K^\pm$ and $B^\pm \to \omega \pi^\pm$ 
by performing 2D $\mb-\Delta E$ fits to the $B^+$ and $B^-$
separately, as shown in Fig.~\ref{acpetap}.
The number of signal events in the $\omega K^\pm$, and $\omega
\pi^\pm$ modes are $21.0^{+6.4}_{-5.7}$, and $10.7^{+6.1}_{-5.2}$
for $B^+$ decays, and $23.6^{+6.6}_{-5.9}$, and
$32.2^{+8.2}_{-7.4}$ for $B^-$ decays, respectively. 
The corresponding partial rate asymmetry values are ${\mathcal A_{CP}} =
0.06^{+0.21}_{-0.18}\pm 0.01$ for $\omega K^\pm$ and ${\mathcal
A_{CP}} = 0.50^{+0.23}_{-0.20}\pm 0.02$ for $\omega \pi^\pm$. The
systematic error in ${\mathcal A_{CP}}$ comes mainly from the reconstruction
efficiency of high momentum charged particles and the fitting
functions. The latter is measured by varying the parameters of the
fitting functions. The asymmetry in $K^{\pm}$ reconstruction
efficiency is studied with an inclusive charged kaon sample.

In the confidence level calculation, we expand the interval
determined solely from the statistical error by one 
standard deviation of the systematic error.
The 90\% confidence
level interval corresponds to $-0.25 < {\mathcal A_{CP}} <0.41$
for $B^{\pm}\to \omega K^{\pm}$ and $0.15 < {\mathcal A_{CP}}
<0.90$ for $B^{\pm} \to \omega \pi^{\pm}$.
\section{Discussion and Conclusion}
In summary, we have searched for exclusive two-body charmless
hadronic $B$ decays with an $\omega$ meson in the final state using
a data sample of 78.1 fb$^{-1}$ collected 
on the $\Upsilon(4S)$ resonance.
We find $\mathcal B$($B^+\to\omega K^+$) =
$(6.5^{+1.3}_{-1.2}\pm {0.6})\times 10^{-6}$ 
and $\mathcal B$($B^+\to\omega \pi^+$) = $(5.7^{+1.4}_{-1.3}\pm 0.6)\times
10^{-6}$, where the first error is statistical and the second
systematic. Our results confirm our previous measurement of a
large branching fractions for $B^+\to \omega K^+$, which cannot be
easily accommodated by the factorization approach and might
indicate the presence of a large non-factorizable contribution or
other penguin related processes~\cite{ntheory}. 
An signal is
obtained for $B^0 \to \omega K^0$ decay with $3.2 \sigma$ significance
while no excess is observed for $B^0 \to \omega \pi^0$
decay. The results correspond to 90\% confidence level upper limits of 
$\mathcal B$($B^0 \to\omega K^0$) $<7.6 \times 10^{-6}$ 
and $\mathcal B$($B^0 \to\omega \pi^0$) $< 1.9 \times 10^{-6}$.

We also search for partial rate asymmetries
in $B^\pm \to \omega K^\pm$ and $\omega \pi^\pm$.
We find ${\mathcal A_{CP}} =
0.06^{+0.21}_{-0.18}\pm 0.01$ for $\omega K^\pm$ and ${\mathcal
A_{CP}} = 0.50^{+0.23}_{-0.20}\pm 0.02$ for $\omega \pi^\pm$. 
These correspond to 90\% confidence 
level intervals of 
$-0.25 < \mathcal A_{cp} < 0.41$ for 
$B^\pm \to \omega K^\pm$ and 
$0.15 < \mathcal A_{cp} < 0.90$ for $B^\pm \to \omega \pi^\pm$.
Our results indicate the possibility of non-zero $\mathcal A_{cp}$
for $B^\pm \to \omega \pi^\pm$ with $99.2\%$ confidence level,
equivalent to $2.4\sigma$ significance for Gaussian errors.
\section{Acknowledgements}
We wish to thank the KEKB accelerator group for the excellent
operation of the KEKB accelerator.
We acknowledge support from the Ministry of Education,
Culture, Sports, Science, and Technology of Japan
and the Japan Society for the Promotion of Science;
the Australian Research Council
and the Australian Department of Education, Science and Training;
the National Science Foundation of China under contract No.~10175071;
the Department of Science and Technology of India;
the BK21 program of the Ministry of Education of Korea
and the CHEP SRC program of the Korea Science and Engineering Foundation;
the Polish State Committee for Scientific Research
under contract No.~2P03B 01324;
the Ministry of Science and Technology of the Russian Federation;
the Ministry of Education, Science and Sport of the Republic of Slovenia;
the National Science Council and the Ministry of Education of Taiwan;
and the U.S.\ Department of Energy.                                             

\begin{thebibliography}{99}
\bibitem{qq}{The QQ B meson decay event generator was developed
by the CLEO Collaboration.}
\end{thebibliography}
\end{document}